\documentclass[prc,twocolumn,showpacs,preprintnumbers,amsmath,amssymb,superscriptaddress,floatfix,nofootinbib]{revtex4}

\usepackage{graphicx}
\usepackage{amsmath}
\usepackage{amsfonts}
\usepackage{amssymb}
\usepackage{color}

\newcommand{\Slash}[1]{\ooalign{\hfil/\hfil\crcr$#1$}}

\begin{document}
\title {Re-analysis of the $\Lambda(1520)$ photoproduction reaction}
\author{Ju-Jun Xie} \email{xiejujun@impcas.ac.cn}
\affiliation{Institute of Modern Physics, Chinese Academy of
Sciences, Lanzhou 730000, China} \affiliation{Instituto de F\'\i
sica Corpuscular (IFIC), Centro Mixto CSIC-Universidad de Valencia,
Institutos de Investigaci\'on de Paterna, Aptd. 22085, E-46071
Valencia, Spain} \affiliation{State Key Laboratory of Theoretical
Physics, Institute of Theoretical Physics, Chinese Academy of
Sciences, Beijing 100190, China}
\author{En Wang}  \email{En.Wang@ific.uv.es}
\affiliation{Instituto de F\'\i sica Corpuscular (IFIC), Centro
Mixto CSIC-Universidad de Valencia, Institutos de Investigaci\'on de
Paterna, Aptd. 22085, E-46071 Valencia, Spain}
\author{J. Nieves}  \email{jmnieves@ific.uv.es}
\affiliation{Instituto de F\'\i sica Corpuscular (IFIC), Centro
Mixto CSIC-Universidad de Valencia, Institutos de Investigaci\'on de
Paterna, Aptd. 22085, E-46071 Valencia, Spain}
\date{\today}%

\begin{abstract}

Based on previous studies that support the important role of the
$N^*(2120) D_{13}$ resonance in the $\gamma p \to K^+ \Lambda(1520)$
reaction, we make a re-analysis of this $\Lambda(1520)$
photoproduction reaction taking into account the recent CLAS
differential cross-section data. In addition to the contact,
$t-$channel $\bar K$ exchange, $s-$channel nucleon pole and
$N^*(2120)$ [previously called $N^*(2080)$] resonance contributions,
which have been already considered in previous works, we also study
the $u-$channel $\Lambda(1115)$ hyperon pole term.  The latter
mechanism has always been ignored in all theoretical analysis, which
has mostly relied on the very forward $K^+$ angular LEPS data. It is
shown that when the contributions from the $N^*(2120)$ resonance and
the $\Lambda(1115)$ hyperon are taken into account, both the new
CLAS and the previous LEPS data can be simultaneously described. We
also show that the contribution from the $u-$channel $\Lambda(1115)$
pole term produces an enhancement for large $K^+$ angles, and it
becomes more and more relevant as the photon energy increases, being
essential to describe the CLAS differential cross sections at
backward angles. Furthermore, we find that the new CLAS data also
favor the existence of the $N^*(2120)$ resonance, and that these
measurements can be used to further constrain its properties.

\end{abstract}
 \maketitle

\section{Introduction}

The baryon spectrum and baryon couplings studied from experimental
data are two of the most important issues in hadronic physics and
they are attracting great attention (see Ref.~\cite{klempt} for a
general review). Nucleon excited states  below $2.0$ GeV have been
extensively studied, from both the experimental and the theoretical
points of view~\cite{pdg2012}. Thus, there exists abundant
information on most of their parameters, such as masses, total and
partial decay widths, and decay modes. However, the current
knowledge on the properties of states around or above $2.0$ GeV is
still in its infancy~\cite{pdg2012}. On the other hand, in this
region of energies, many theoretical predicted {\it missing $N^*$
states}, within the constituent quark~\cite{capstick2000} or chiral
unitary~\cite{Sarkar:2009kx,
Oset:2009vf,Gamermann:2011mq,PavonValderrama:2011gp,Sun:2011fr}
approaches, have so far not been observed. Because a large number of
effective degrees of freedom will induce a great number of excited
states, the {\it missing $N^*$ states} problem  seems to favor
diquark configurations, which could lead to reduced numbers of
degrees of freedom. Such schemes would naturally predict a smaller
number of excited $N^*$ states~\cite{Zou:2008be}. Thus, the study of
the possible role played by the $2.0$ GeV region nucleon resonances
in the available new accurate data from the LEPS and CLAS
Collaborations is timely and could shed light on the complicated
dynamics that governs the highly excited nucleon (or in general
baryon) spectrum.

The associated strangeness production reaction $\gamma p \to K^+
\Lambda(1520)$ might be adequate to study the $N^*$ resonances
around $2.0$ GeV, as long as they have significant couplings to the
$K\Lambda(1520)$ pair. This is because the $K\Lambda(1520)$ is a
pure isospin $1/2$ channel and the threshold is about $2.0$ GeV
($m_K + M_{\Lambda(1520)} \simeq 2.0$ GeV). Besides, this reaction
requires the creation of an $\bar{s}s$ quark pair. Thus, a thorough
and dedicated study of the strangeness production mechanism in this
reaction has the potential to achieve a deeper understanding of the
interaction among strange hadrons and, also, of the nature of the
baryon resonances.

There were pioneering measurements at Cornell~\cite{cornell} and
CEA~\cite{CEA}, and in the 1970s, the first $\gamma p \to K^+
\Lambda(1520)$ cross sections in the high-energy region $E_\gamma$ =
11 GeV (SLAC~\cite{boyarski71}), and in the range $2.8$-$4.8$ GeV
(LAMP2 Collaboration~\cite{barber80}) were reported. In 2001, the
CLAS Collaboration investigated this process in
electroproduction~\cite{Barrow:2001ds}, at electron beam energies of
4.05, 4.25, and 4.46 GeV, in the kinematic region spanning the
squared momentum transfer $Q^2$ from 0.9 to 2.4 GeV$^2$, and for
invariant masses from 1.95 to 2.65 GeV. Later, in 2010, this
reaction was examined at photon energies below 2.4 GeV in the
SPring-8 LEPS experiment using a forward-angle spectrometer and
polarized photons~\cite{leps1,leps2}, and from threshold to 2.65 GeV
with the SAPHIR detector at the electron stretcher facility ELSA in
Bonn~\cite{Wieland:2011zz}. Very recently, the exclusive
$\Lambda(1520)$ ($\equiv \Lambda^*$) photoproduction  cross section
has been measured by using the CLAS detector for energies from
threshold up to an invariant $\gamma p$ mass $W = 2.85$ GeV
~\cite{Moriya:2013hwg}.

The theoretical activity has run in parallel.  There exist several
effective hadron Lagrangian
studies~\cite{nam1,nam,nam2,titovprc7274,toki,xiejuan,hejun} exist
for laboratory photon energies ranging from threshold up to about 5
GeV. These theoretical studies have traditionally been limited by
the lack of knowledge on the $\bar K^* N \Lambda^*$ coupling
strength. This fact, in conjunction with the use of largely
different form factors to account for the compositeness of the
hadrons, has led to contradicting predictions of the dominant
reaction mechanism in the process.  Some light was shed on this
issue in \cite{toki}. There, the SU(6) Weinberg-Tomozawa chiral
unitary model proposed in \cite{GarciaRecio:2005hy} was used to
predict a relatively small $\bar K^* N\Lambda^*$ coupling and,
hence, to conclude that the $\bar K$ exchange and contact mechanisms
dominated the $\gamma p \to K^+ \Lambda(1520)$ reaction. Besides, in
the higher energy region, the quark-gluon string mechanism with the
$\bar K$ Regge trajectory was shown~\cite{toki} to reproduce both
the LAMP2 energy and the angular distribution data~\cite{barber80}.

The theory groups have also paid attention to another distinctive
feature of the data. The LEPS energy dependence of the forward-angle
cross section rises from threshold to a maximum near $W$ = 2.15 GeV,
followed by a decline~\cite{leps2}. It was suggested that this could
be an effect of the odd parity $D_{13}$ ($L_{2T2J}$) $N^*$
intermediate resonance at around 2.1 GeV~\cite{nam2,
xiejuan,hejun,nam3}. In the scheme of Refs.~\cite{nam2}
and~\cite{nam3}, the contribution of the spin-parity $J^P=3/2^-$
$N^*(2080)$ ($\equiv N^*$) resonance\footnote{Before the 2012
Particle Date Group (PDG) review, all the evidence for a $J^P=3/2^-$
state with a mass above 1.8 GeV was filed under a two-star
$N^*(2080)$. There is now evidence~\cite{Anisovich:2011fc} of two
states in this region, and the PDG has split the older data
(according to mass) between a three-star $N^*(1875)$ and a two-star
$N^*(2120)$~\cite{pdg2012}.} turned out to be very small. Hence in
these works, such a contribution was not expected to explain the
bump structure at forward angles reported by LEPS. This was
primarily owing to the unnecessarily small $N^*\Lambda^* K$ coupling
and probably the excessively large width of the resonance used in
these references. However in Ref.~\cite{xiejuan}, within the
effective Lagrangian approach of Ref.~\cite{toki}, the role played
by the $N^*(2120)$ resonance in the $\gamma p \to K^+ \Lambda(1520)$
reaction was revisited, and found that the experimental LEPS
Collaboration data could be fairly well described assuming a large
$N^*\Lambda^* K$ coupling, which would be supported by some
constituent quark models~\cite{simonprd58}. Yet, the recent analysis
carried out in \cite{Nam:2012ui} taking the hadron resonance
couplings from ~\cite{simonprd58}, shows that the $D_{13}$
$N^*(2120)$ resonance also plays an important role in reproducing
the $\Lambda(1520)$ electroproduction CLAS data~\cite{Barrow:2001ds}
properly. A large $N^*\Lambda^* K$ coupling scenario has also been
investigated in the $pp \to p K^+ \Lambda^*$ and $\pi^- p \to K^0
\Lambda^*$ hadronic $\Lambda(1520)$ production
reactions~\cite{xieliu}. Note that a resonance with these quantum
numbers in the 2.1 GeV energy region, albeit with some uncertainty
in the precise position, is unavoidable owing to the attractive
character and strength of the vector-baryon interaction within the
schemes of Refs.~\cite{Gamermann:2011mq} and \cite{Oset:2009vf}.
Furthermore, a recent analysis~\cite{Ramos:2013wua} of the $\gamma p
\to K^0\Sigma^+$ CBELSA/TAPS data \cite{TAPS}, which exhibits a peak
in the cross section around $\sqrt{s}$ = 1.9 GeV followed by a fast
downfall around $\sqrt{s}$ = 2.0 GeV, also provides support for the
existence of a $J^P=3/2^-$ nucleon excited state around 2 GeV.

On the other hand, the comparison of the recent CLAS
measurements~\cite{Moriya:2013hwg} with different effective
Lagrangian model predictions~\cite{hejun,nam}, obtained by using the
parameters fitted to the LEPS~\cite{leps2,leps1} and
LAMP2~\cite{barber80} data, indicates that the current model
calculations can not describe the CLAS differential cross sections
well over the entire energy and angular ranges available in the
experiment.

In the present work, we aim to achieve an improved description of
the recent CLAS data, which would provide further support for the
existence of the $N^*(2120)$ resonance, and additional constraints
on its properties. Within the scheme of Ref.~\cite{xiejuan}, and in
addition to the contact, $t-$channel $\bar K$ exchange, and
$s-$channel nucleon  and $N^*(2120)$ resonance pole contributions,
we also study the $u-$channel $\Lambda(1115)$ ($\equiv \Lambda$)
hyperon pole term. The latter mechanism has been ignored in all
previous calculations~\cite{xiejuan,hejun,nam,toki} because
information on the $\Lambda^* \Lambda \gamma$ coupling is
scarce~\cite{titovprc7274}, and because it is expected to provide
small contributions for forward $K^+$ angles, where the LEPS data
lie. However, $u-$channel mechanisms might make important
contributions at the backward $K^+$ angles that have become
accessible in the recent CLAS experiment. As a consequence, it is
not surprising, that previous theoretical calculations that
reasonably described the LEPS data, were not as successful in
describing the recent CLAS differential cross sections, which span a
much wider range of $K^+$ angles.

Taking these considerations into account, and using the combination
of the effective Lagrangian approach and the isobar model, we
present in this work a combined  theoretical analysis of the recent
$\gamma p \to K^+ \Lambda(1520)$  CLAS~\cite{Moriya:2013hwg} and
LEPS~\cite{leps2} data that includes the contribution from the
$u-$channel $\Lambda$ hyperon pole term.

The paper is organized as follows. In Sec.~\ref{sec:formalism}, we
discuss the formalism and the main ingredients of the model, while
our numerical results and discussion are presented in
Sec.~\ref{sec:results}. Finally, a short summary and conclusions are
given in  Sec.~\ref{sec:conclusions}.
%
%%%%%%%%%%%%%%%%%%%%%%%%%%%%%%%%%%%%%%%%%%%%%%%%%%%%%%%%%%%%%%%%%%
%
\section{Formalism and ingredients} \label{sec:formalism}
\begin{figure}[htdp]
\includegraphics[scale=0.4]{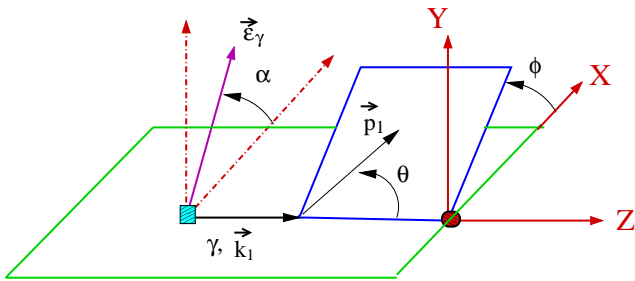}
\vspace{-0.2cm} \caption{ (Color online) Definition of the different
angles used in this work.} \label{plane}
\end{figure}
\begin{figure}[htdp]
\includegraphics[scale=0.8]{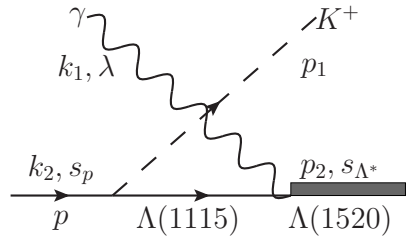}
\caption{The $u-$channel $\Lambda$ pole mechanism. In the diagram,
we also show the definition of the kinematical $(k_1, k_2, p_1,p_2)$
and polarization $(\lambda, s_p, s_{\Lambda^*})$ variables that we
use in the present calculation.  In addition, we use $q_u =
k_2-p_1=p_2-k_1$, with $u=q_u^2$.} \label{fig:uterm}
\end{figure}

The invariant scattering amplitudes that enter our model for
calculation of the total and differential cross sections for the
reaction\footnote{The definition of the kinematical and polarization
variables is the same as in Ref.~\cite{xiejuan} [see also
Eq.~(\ref{eq:reac}) and Figs.~\ref{plane} and ~\ref{fig:uterm}].}
\begin{equation}
\gamma(k_1,\lambda) p(k_2,s_p) \to K^+ (p_1)  \Lambda^* (p_2,s_{\Lambda^*}) \label{eq:reac}
\end{equation}
are defined as
\begin{equation}
-iT_i=\bar u_\mu(p_2,s_{\Lambda^*}) A_i^{\mu \nu} u(k_2,s_p)
\epsilon_\nu(k_1,\lambda)
\end{equation}
where $u_\mu$ and $u$ are dimensionless Rarita-Schwinger and Dirac
spinors, respectively, while $\epsilon_\nu(k_1,\lambda)$ is the
photon polarization vector. In addition, $s_p$ and $s_{\Lambda^*}$
are the baryon polarization variables. The sub-index $i$ stands for
the contact, $t-$channel antikaon exchange, $s-$channel nucleon and
$N^*$ pole terms (depicted in Fig.~1 of Ref.~\cite{xiejuan}) and
novel $u-$channel $\Lambda$ pole mechanism. In our final results,
and for simplicity, we do not consider the $u-$channel
$\Sigma^0(1193)$ pole term, as we expect its contribution to be much
smaller than that from the $\Lambda (1115)$ hyperon. We come back to
this point below. Moreover, higher-excited hyperons would be farther
off-shell in the $u-$channel, and have not been taken into account
either.

The explicit expressions for the  reduced $A_i^{\mu\nu}$ amplitudes
of the first four mechanisms, Fig.~1 of Ref.~\cite{xiejuan}, can be
found in that reference. Here, we only give details on the
$u-$channel $\Lambda$ pole amplitude, $A^{\mu\nu}_u$, associated
with the diagram in Fig.~\ref{fig:uterm}. It is obtained from the
effective interaction Lagrangian densities,
\begin{eqnarray}
\mathcal{L}_{\gamma \Lambda \Lambda^*}  &=&
-\frac{ih_1}{2m_{\Lambda}} \bar{\Lambda}^*_{\mu} \gamma_{\nu} F^{\mu
\nu} \Lambda \nonumber \\
&&+\frac{h_2}{(2m_{\Lambda})^2}\bar{\Lambda}^*_{\mu} F^{\mu \nu}
\partial_{\nu} \Lambda\,+{\rm h.c.} \\ \label{eq:eqgamaN}
\mathcal{L}_{K N \Lambda}   &=&  -i g_{KN\Lambda}
\bar{\Lambda} \gamma_5 K^- p \,+{\rm h.c.}, \label{eq:eqknstar}
\end{eqnarray}
Note that the $\gamma\Lambda\Lambda^*$ vertex is gauge invariant by
itself, as $A_\mu$ and $F_{\mu\nu} = (\partial_\mu A_\nu -
\partial_\nu A_\mu)$ are the photon field and
electromagnetic-field tensors, respectively. $h_1$ and $h_2$ are
magnetic coupling constants, while $g_{KN\Lambda}$ is a strong one.
We take $g_{KN\Lambda} \sim -14$ as  estimated from the SU(3) flavor
symmetry~\cite{swart6365} Bonn-J\"ulich model for the meson-exchange
hyperon-nucleon interactions in free
scattering~\cite{Reuber:1993ip}. Note that at lowest order in the
chiral expansion~\cite{Pich,Scherer:2012xha},
\begin{equation}
\frac{g_{\pi^0pp}}{g_{K^+p\Lambda}} \sim -
\frac{D+F}{\frac{D+3F}{\sqrt{3}}}, \qquad
\frac{g_{K^+p\Sigma^0}}{g_{K^+p\Lambda}} \sim -
\frac{D-F}{\frac{D+3F}{\sqrt{3}}}\label{eq:chiral},
\end{equation}
with $D\sim 0.8$ and $F\sim 0.5$, which justifies
$|g_{KN\Lambda}/g_{\pi NN}|\sim 1$. Besides, we see that the
$g_{K^+p\Sigma^0}$ is about four or five times smaller than
$g_{K^+p\Lambda}$, and this is also compatible with the findings in
~\cite{Reuber:1993ip}. The latter relationship between couplings
gives support for neglecting the $u-$channel $\Sigma^0$ pole
contribution, despite that the strength of the electromagnetic
$\gamma\Sigma^0 \Lambda^*$ coupling could be similar to, or even
larger than, that involving the $\Lambda(1115)$
hyperon~\cite{Mast:1969tx}.
The $\Sigma^0$ radiative decay of the $\Lambda(1520)$ has not been
observed yet, while $\Gamma(\Lambda(1520)\to \Lambda\gamma)$ has
been measured to be of the order of 130 keV \cite{pdg2012}. There
exist various (Table~\ref{tab:sigma}) theoretical
estimates~\cite{DHK83,Kaxiras:1985zv,Warns:1990xi, Umino:1991dk,
Bijker:2000gq,VanCauteren:2005sm, Yu:2006sc, Doring:2006ub} for the
$\Gamma(\Lambda(1520)\to \Sigma^0\gamma)$ width, ranging from the 17
keV predicted within the MIT bag model~\cite{Kaxiras:1985zv} up to
the 293 keV obtained in the RCQM (relativistic constituent quark
model) approach followed in \cite{Warns:1990xi}. If the
$\Lambda(1520)$ were a SU(3) singlet and assuming that the photon is
a $U-$spin singlet, $U-$spin invariance guarantees that decay into
the $U-$spin triplet linear combination $(\Sigma^0+3\Lambda)$ would
be 0~\cite{Mast:1969tx}. Thus one would predict\footnote{In the
original work in Ref.~\cite{Mast:1969tx}, it was assumed that the
phase space scaled as the decaying photon momentum from what follows
$\Gamma(\Lambda(1520)\to \Sigma^0\gamma) \sim 2.5 \times
\Gamma(\Lambda(1520)\to\Lambda\gamma)$~\cite{Mast:1969tx}. However,
as we see below (Eqs.~(\ref{eq:Gammai})--(\ref{eq:Gammaf})), the
decay width rather scales down as the cube of the photon momentum,
which leads to the lower value, about 230 keV.}
$\Gamma(\Lambda(1520)\to \Sigma^0\gamma) \sim 3\times$ (phase space)
$\times \Gamma(\Lambda(1520)\to \Lambda\gamma)\sim 230$ keV, in the
upper-band of the various predictions listed in
Table~\ref{tab:sigma}.

\begin{table*}
\caption{Theoretical predictions (in
  keV) for the radiative width $\Gamma(\Lambda(1520)\to
  \Sigma^0\gamma)$.}
\vspace{0.4cm}
\begin{tabular}{ccccccccc}
\hline \hline
Approach & NRQM & MIT Bag & RCQM & Chiral Bag & Algebraic & BonnBSCQM &$\chi$QM & UChPT \\
Ref. & \cite{DHK83, Kaxiras:1985zv} & \cite{Kaxiras:1985zv} &
\cite{Warns:1990xi} & \cite{Umino:1991dk} & \cite{Bijker:2000gq} &
\cite{VanCauteren:2005sm} &
\cite{Yu:2006sc} & \cite{Doring:2006ub}\\
$\Gamma$ [keV] & 55--75 & 17 & 293 & 49 & 180 & 157 & 92 & 71 \\
\hline \hline
\end{tabular}\label{tab:sigma}
\end{table*}

Despite this large ambiguity, one can safely conclude that even
though the electromagnetic $\gamma\Sigma^0 \Lambda^*$ coupling was
bigger than the $\gamma\Lambda \Lambda^*$ one, it would not be large
enough to compensate for the big reduction induced by the ratio of
hadronic couplings $\left(g_{K^+p\Sigma^0}/g_{K^+p\Lambda}\right)^2
\sim 1/20$. Hence, it seems reasonable to assume that the
$\Sigma^0(1193)$ contribution to the $\gamma p \to K^+
\Lambda(1520)$ reaction would be much smaller than that of the
$\Lambda (1115)$ hyperon. On the other hand, taking into account the
$u-$channel $\Sigma^0(1193)$ term would require adopting some
theoretical model for the electromagnetic $\Lambda^*\Sigma^0\gamma$
amplitude. However, we see that important discrepancies among the
various approaches, even in predicting the radiative decay width
$\Gamma(\Lambda(1520)\to \Sigma^0\gamma)$. Fitting this amplitude to
data is not a realistic option either, because both the $\Lambda$
and the $\Sigma^0$ $u-$channel poles are quite close, and the
available data cannot effectively discriminate between their
contributions. Thus, in the final results, and for the sake of
simplicity, we do not include the $u-$channel $\Sigma^0(1193)$
mechanism. Nevertheless, we estimate the possible impact of this
term assuming $U-$spin SU(3) symmetry.

With the above Lagrangians, one readily finds
\begin{eqnarray}
A_{u}^{\mu\nu} \!\!\!\! &=&  \!\!\!\!  \Big
(\frac{h_1}{2m_{\Lambda}}(k_1^{\mu}\gamma^{\nu}- \!\!
g^{\mu\nu}\Slash k_1) + \!\! \frac{h_2}{(2m_{\Lambda})^2}
(k_1^{\mu}q_u^{\nu}- \!\! g^{\mu\nu}k_1 \cdot q_u)\Big) \nonumber \\
&&  \times \frac{\Slash q_u + m_{\Lambda}}{u - m_{\Lambda}^2}
g_{KN\Lambda}
 \gamma_5 f_u, \label{eq:au}
\end{eqnarray}
where $m_{\Lambda}$ and $q_u$ are the mass and the four-momentum of
the $\Lambda (1115)$, respectively, and $u = q^2_u$. In addition, we
should introduce in the above equation the compositeness of the
hadrons. This has been achieved by including the form factor $f_u$
in the amplitude of Eq.~(\ref{eq:au}),
\begin{eqnarray}
f_{u} &=&\frac{\Lambda^4_u}{\Lambda^4_u+(u - m_{\Lambda}^2)^2},
\end{eqnarray}
whose form is similar to those used in ~\cite{xiejuan} to construct
the form-factors that appear in the rest of the amplitudes in the
model. In principle, the cut off $\Lambda_u$ is a free parameter of
the model, but in practice we will fix it to a common value taken by
$\Lambda_t$ and $\Lambda_s$ as well.\footnote{These are other cut
offs that enter into the form factors that we include in the
$t-$channel antikaon exchange, $s-$channel nucleon pole and contact
amplitudes~\cite{xiejuan}.}

The $ \gamma\Lambda \Lambda^*$ coupling constants $h_1$ and $h_2$
could be fixed  from the $\Lambda^* \to \Lambda\gamma$
partial decay width~\cite{pdg2012},
\begin{eqnarray}
\Gamma_{\gamma} = \frac{m_{\Lambda}k^2_{\gamma}}{2\pi M_{\Lambda^*}}
\left(|A_{1/2}|^2 + |A_{3/2}|^2\right)\label{eq:Gammai},
\end{eqnarray}
with,
\begin{eqnarray}
A_{1/2} \!\!\! &=& \!\!\!
\frac{\sqrt{6}}{12}\sqrt{\frac{k_{\gamma}}{m_{\Lambda}
M_{\Lambda^*}}}\left( h_1+\frac{M_{\Lambda^*}h_2}{4m^2_{\Lambda}}(M_{\Lambda^*}+m_{\Lambda}) \right ) \\
A_{3/2} \!\!\! &=& \!\!\!
\frac{\sqrt{2}}{4m_{\Lambda}}\sqrt{\frac{k_{\gamma}
M_{\Lambda^*}}{m_{\Lambda}}} \left(
h_1+\frac{h_2}{4m_{\Lambda}}(M_{\Lambda^*}+m_{\Lambda}) \right ),\label{eq:Gammaf}
\end{eqnarray}
where $k_{\gamma} =
(M^2_{\Lambda^*}-m^2_{\Lambda})/(2M_{\Lambda^*})$ is the photon
 center of mass (c.m.) frame  decay momentum. With the value of
$\Gamma_{\gamma} = 0.133$ MeV, as quoted in the PDG~\cite{pdg2012},
we could get a constraint on the values of $h_1$ and $h_2$,
\begin{eqnarray}
ah^2_1+bh^2_2+ch_1h_2=d, \label{eq:h1h2}
\end{eqnarray}
with
\begin{eqnarray}
a \!\!\! &=& \!\!\!\!
\frac{m^2_{\Lambda}+3M^2_{\Lambda^*}}{24m^3_{\Lambda}M_{\Lambda^*}}k^3_{\gamma},~~~~
b =
\frac{M_{\Lambda^*}(m_{\Lambda}+M_{\Lambda^*})^2}{96m^5_{\Lambda}}k^3_{\gamma},
\\
c  \!\!\! &=& \!\!\!\!
\frac{(m_{\Lambda}+M_{\Lambda^*})(m_{\Lambda}+3M_{\Lambda^*})}{48m^4_{\Lambda}}k^3_{\gamma},~~
d \! = \!\! \frac{2\pi M_{\Lambda^*} \Gamma_{\gamma}}{m_{\Lambda}}
\end{eqnarray}

With the ingredients given above, the unpolarized
c.m. differential cross section  can be easily obtained as,
\begin{eqnarray}
\frac{d\sigma}{d(\cos\theta_{\rm c.m.})}  &=&
\frac{|\vec{k}_1^{\rm \,\, c.m.}||\vec{p}_1^{\rm
\,\,c.m.}|}{4\pi}\frac{M_N M_{\Lambda^*}}{(W^2 - M_N^2)^2} \nonumber \\
&\times& \sum_{s_p, s_{\Lambda^*},\lambda} |T|^2
 \label{eqdcs}
\end{eqnarray}
with $W$ the invariant mass of the $\gamma p$ pair. Further,
$\vec{k}_1^{\rm \,\, c.m.}$ and $\vec{p}_1^{\rm \,\, c.m.}$ are the
photon and $K^+$ meson c.m. three-momenta, and $\theta_{\rm c.m.}$
is the  $K^+$ polar scattering angle (Fig.~\ref{plane}). The
differential cross section $d\sigma/d(\cos\theta_{\rm c.m.})$
depends on $W$ and also on $\cos\theta_{\rm c.m.}$.

As mentioned above, the model accounts for a total of five
mechanisms: contact, $t-$channel antikaon exchange, $s-$channel
nucleon and $N^*$ pole terms, evaluated in ~\cite{xiejuan}, and the
$u-$channel $\Lambda$ pole contribution discussed here. In
principle, the free parameters of the model are: (i)  the mass and
width ($M_{N^*}$ and $\Gamma_{N^*}$) of the $N^*(2120)$ resonance,
(ii) the cut off parameters $\Lambda_s =\Lambda_t = \Lambda_u \equiv
\Lambda_B$ and  $\Lambda_R$, and (iii) the $N^*(2120)$ resonance
electromagnetic $\gamma NN^*$ ($ef_1$, $ef_2$) and strong
$N^*\Lambda^*K$ ($g_1$, $g_2$) couplings and the $\Lambda(1520)$
magnetic $\gamma\Lambda\Lambda^*$ ($h_1$) one. Note that the second
coupling $h_2$ in Eq.~(\ref{eq:eqgamaN}) is given in terms of $h_1$
and the $\Lambda^*$ radiative decay width (see Eq.~(\ref{eq:h1h2})).

In the next section, we fit the parameters of the model to the
differential cross-section data from the CLAS and LEPS
Collaborations.

\section{Numerical results and discussion} \label{sec:results}

%%%
\begin{table*}
\caption{Values of some parameters determined in this work and in
Ref.~\cite{xiejuan}. Fit I (II) parameters have been adjusted to the
CLAS (combined CLAS~\cite{Moriya:2013hwg} and LEPS~\cite{leps2})
$\gamma p \to K^+ \Lambda(1520)$ $d\sigma/d(\cos\theta_{\rm c.m.})$
data, while fit C in Ref.~\cite{xiejuan} was obtained by considering
only the LEPS differential cross sections. LEPS data lie in the
$K^+$ forward angle region and were taken below $E_\gamma$ = 2.4
GeV, while the recent CLAS measurements span a much larger $K^+$
angular and photon energy regions. Finally, we also list for each
fit the predicted $N^*(2120)$ width $\Gamma_{N^* \to \Lambda^*K}$,
helicity amplitudes for the positive-charge state and the $h_2$
electromagnetic $\gamma\Lambda\Lambda^*$ coupling.} \vspace{0.4cm}
\begin{tabular}{cccc}
\hline \hline & \multicolumn{3}{c}{Fitted Parameters} \\
 & \multicolumn{2}{c}{This work}& Ref.~\cite{xiejuan} \\
 & Fit I & Fit II & Fit C \\
        $g_1$ & $1.7 \pm 0.4$ & $1.6 \pm 0.2$   & $1.4 \pm 0.3$ \\
        $g_2$ & $4.6 \pm 1.2$ & $2.2 \pm 0.5$   & $5.5 \pm 1.8$ \\
        $\Lambda_B$ [MeV] & $630 \pm 2$ & $620 \pm 2$    & $604 \pm 2$ \\
        $\Lambda_R$ [MeV] & $933 \pm 52$ & $1154 \pm 47$  & $909 \pm 55$ \\
        $ef_1$ & $0.123 \pm 0.015$ & $0.126 \pm 0.012$    & $0.177 \pm 0.023$ \\
        $ef_2$  & $-0.094 \pm 0.014$ & $-0.097 \pm 0.010$   & $-0.082 \pm 0.023$ \\
        $M_{N^*}$[MeV] & $2172 \pm 10$ & $2135 \pm 4$   & $2115 \pm 8$ \\
        $\Gamma_{N^*}$[MeV] & $287 \pm 54$ & $184 \pm 11$   & $254 \pm 24$ \\
        $h_1$ & $0.68 \pm 0.05$ & $0.64 \pm 0.05$ & $-$ \\
        $\chi^2/dof$ & $2.5$ & $2.5$ & $1.2$ \\
        $\chi^2/dof$ (without $N^*$) & $5.6$ & $9.9$ & $24$ \\
        $\chi^2/dof$ (without $\Lambda$) & $3.0$ & $3.0$ & $-$ \\
        $\chi^2/dof$ (without $N^*$ and $\Lambda$) & $6.3$ & $9.9$ & $-$ \\
\hline
\hline  & \multicolumn{3}{c}{Derived
  Observables}  \\
  $A_{1/2}^{p^*}[10^{-3}\text{GeV}^{-1/2}]$ & $-7.6 \pm 3.9$ & $-7.3 \pm 3.0$  & $3.6 \pm 8.6$  \\
        $A_{3/2}^{p^*}[10^{-2}\text{GeV}^{-1/2}]$ & $2.5 \pm 1.0$ & $2.5 \pm 0.8$   & $5.8 \pm 2.1$ \\
        $\Gamma_{N^* \to \Lambda^*K}$ [MeV] & $56 \pm 27$ & $30 \pm 8 $   & $19 \pm 7$ \\
        $\frac{\Gamma_{N^* \to \Lambda^*K}}{\Gamma_{N^*}}$[$\%$] & $19.0 \pm 10.3$   & $16.2 \pm 4.2 $ & $7.5 \pm 2.8$ \\
       $h_2$ & $-0.43 \pm 0.08$ & $-0.38 \pm 0.07$ & $-$ \\
\hline \hline
\end{tabular} \label{tab:nstar}
\end{table*}

\begin{figure*}
\begin{center}
\includegraphics[scale=1.6]{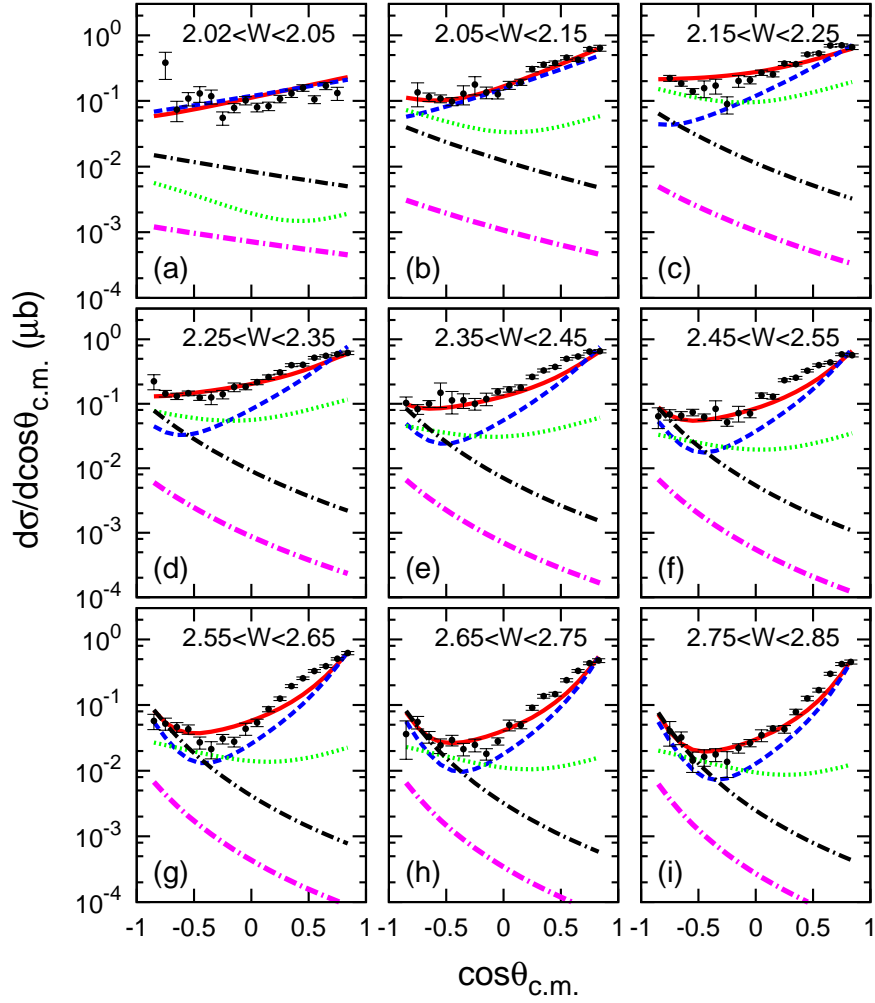}
\caption{(Color online) Fit II $\gamma p \to K^+ \Lambda(1520)$
differential cross sections as a function of $\cos\theta_{\rm c.m.}$
compared with the CLAS data~\cite{Moriya:2013hwg} for different
$\gamma p$-invariant mass intervals (in GeV). Only statistical
errors are displayed. Dashed (blue) and dotted (green) lines show
the contributions from background (contact, $t-$channel $\bar K$
exchange and $s-$channel nucleon and $u-$channel $\Lambda$ pole
mechanisms) and $N^*(2120)$ resonance terms, respectively. The
Dash-dotted (black) curves represent for the $u-$channel $\Lambda
(1115)$ contribution separately, while the solid (red) lines display
the results obtained from the full model (background+$N^*$).
Finally, the dash-dotted (magenta) curves represent the $u-$channel
$\Sigma^0(1193)$ contribution, not included in the final results
[solid (red) lines], as determined from electromagnetic $U-$spin
($h_{1,2}^\Sigma=-\sqrt{3}h_{1,2}$) and leading order chiral
[Eq.~(\ref{eq:chiral})] symmetries.}
\label{dcs-clas}%
\end{center}
\end{figure*}

\begin{figure}
\begin{center}
\includegraphics[scale=1.]{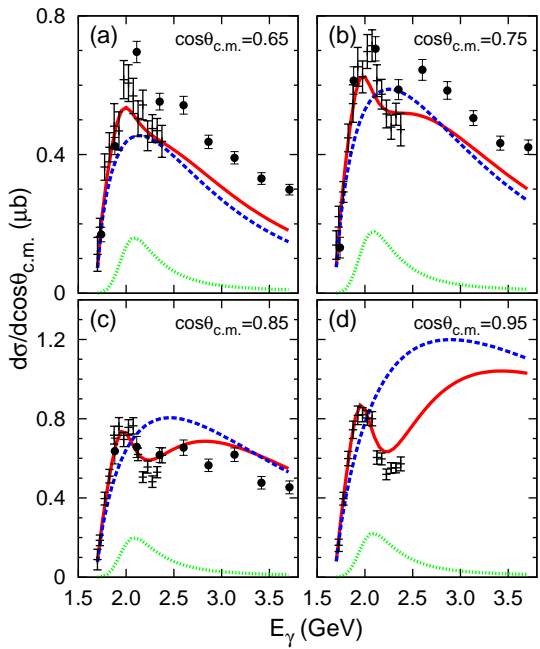}
\caption{(Color online) Fit II $\gamma p \to K^+ \Lambda (1520)$
differential cross section as a function of the LAB frame photon
energy for different c.m. $K^+$ polar angles. We also show the
experimental LEPS~\cite{leps2} (crosses) and
CLAS~\cite{Moriya:2013hwg} (filled circles) data. Only statistical
errors are displayed. Dashed (blue) and dotted (green) lines show
the contributions from the background and $N^*(2120)$ resonance
terms, respectively, while the red-solid lines represent the results
obtained from the full model.}
\label{dcs-leps}%
\end{center}
\end{figure}

First, we have performed\footnote{We take $M_{\Lambda^*}= 1.5195$
  GeV, $m_K$= 0.4937 GeV and $m_{\Lambda}= 1.1157$ GeV.} a nine-parameter
($ef_1$, $ef_2$, $g_1$, $g_2$, $\Lambda_s =\Lambda_t = \Lambda_u
\equiv \Lambda_B$, $\Lambda_R$, $M_{N^*}$, $\Gamma_{N^*}$ and $h_1$)
$\chi^2-$fit to the $d\sigma/d(\cos\theta_{\rm c.m.})$ data from the
CLAS Collaboration~\cite{Moriya:2013hwg} (fit I). There is a total
of $157$ available data points displayed in Fig.~\ref{dcs-clas}. The
$d\sigma/d(\cos\theta_{\rm c.m.})$ data, as a function of
$\cos\theta_{\rm c.m.}$, are given for nine intervals of the
invariant $\gamma p$ mass $W$ from the reaction threshold $2.02$ GeV
up to $2.85$ GeV. To compute the cross sections in each interval we
always use the corresponding mean value of $W$. (We have checked
that variations with respect to calculation of the average value of
the differential cross sections for each $W$ range turn out to be
very small.) We have also carried out a combined fit to the
CLAS~\cite{Moriya:2013hwg} and LEPS~\cite{leps2},
$d\sigma/d(\cos\theta_{\rm c.m.})$ data (fit II). In this second
fit, we have a total of $216$ data points (in addition to the former
CLAS differential cross sections, we have also fitted to the LEPS
data depicted in Fig.~\ref{dcs-leps}).

The fitted parameters from the above two fits are listed in
Table~\ref{tab:nstar}, where we also report our previous results
from a best fit (fit C of Ref.~\cite{xiejuan}) only to the LEPS data
of Ref.~\cite{leps2}. We also give for each fit, the predicted
nucleon $D_{13}$ resonance width $\Gamma_{N^* \to \Lambda^*K}$
(Eq.~(18) in Ref.~\cite{xiejuan}), helicity amplitudes (Eqs.~(15)
and (16) in Ref.~\cite{xiejuan}) for the positive-charge state and
the $h_2$ magnetic $\gamma\Lambda\Lambda^* $ coupling
[Eq.~(\ref{eq:h1h2})].\footnote{This is a second rank equation and
it has two possible solutions. One of them turns out to be strongly
disfavored by the $\chi^2-$fit. To be more quantitative, in our
final results (fit II), $\chi^2/dof$ would pass from 2.5 to 3.1 if
the other solution were considered.}

As commented before, LEPS data lie in the $K^+$ forward angle region
and were taken below $E_\gamma$ = 2.4 GeV, while the recent CLAS
measurements span a much larger $K^+$ angular and photon energy
regions. The $\chi^2/dof$ for both fit I and fit II are acceptable,
of the order of 2.5.  The new CLAS measurements are quite accurate
(excluding some data close to threshold) with statistical errors
ranging from about 10\%, at high scattering angles, down to 5\% or
less for forward angles. The systematical errors of the experimental
data ($11.6\%$~\cite{Moriya:2013hwg} and $5.92\%$~\cite{leps2}, for
CLAS and LEPS, respectively) have been added in quadratures to the
statistical ones and taken into account in the present new fits. We
see that the $N^*$ resonance parameters from the new fits I and II
turn out to be in reasonable agreement with those obtained in
Ref.~\cite{xiejuan}. Thus, the first conclusion is that the CLAS
data provide further support for the existence of an odd parity
$3/2$ wide nucleon resonance with a mass in the region of 2.1 GeV
and a width of around 200 MeV. This is compatible with the
Breit-Wigner parameters, $M_{BW}=2.15 \pm 0.06 $ GeV and
$\Gamma_{BW}=330 \pm 45 $ MeV, reported in \cite{Anisovich:2011fc}.
The latter reference also provides experimental values for its
helicity amplitudes
\begin{eqnarray}
A_{1/2}^{p^*}[10^{-3}\text{GeV}^{-1/2}] &=& 125 \pm 45 \\
A_{3/2}^{p^*}[10^{-2}\text{GeV}^{-1/2}] &=& 15 \pm 6\,,
\end{eqnarray}
which however do not seem entirely
consistent with previous measurements~\cite{Awaji:1981zj}
\begin{eqnarray}
A_{1/2}^{p^*}[10^{-3}\text{GeV}^{-1/2}] &=& -20 \pm 8 \\
A_{3/2}^{p^*}[10^{-2}\text{GeV}^{-1/2}] &=&  1.7 \pm 1.1
\end{eqnarray}
quoted in the 2008 PDG edition~\cite{Amsler:2008zzb}, which in turn
are in better agreement with our predictions in
Table~\ref{tab:nstar}. Note that the latter helicity amplitudes are
used in \cite{Nam:2012ui}, where the $ep \to e K^+ \Lambda(1520)$
CLAS data in Ref.~\cite{Barrow:2001ds} were successfully described.
Nevertheless, given the two-star status (evidence of existence is
only fair) granted to the $N^*(2120)$ resonance in the multichannel
partial wave analysis of pion and photo-induced reactions off
protons carried out in \cite{Anisovich:2011fc}, the discrepancy with
our predicted helicity amplitudes should not be used to rule out our
fits, but rather one could use them to further constrain these
elusive observables. On the other hand, within this scheme, the
$N^*(2120)$ resonance would have a large partial decay width into
$\Lambda^*K$, which is compatible with the findings of the
constituent quark model approach in Ref.~\cite{simonprd58}. Indeed,
in that reference, $\Gamma_{N^* \to \Lambda^*K}$ is predicted to be
$ 7^{+24}_{-6}$ MeV for a resonance mass of 2080 MeV. This value for
the width is compatible within errors with the value of $30\pm 8$
MeV found in this work. Moreover, because the $\Lambda(1520) K^+$
threshold is located so close to $M_{N^*}=2080$ MeV, the width would
increase by at least a factor of two if the resonance mass was
instead taken as 2120 MeV. Hence, the constituent quark model of
Ref.~\cite{simonprd58} would predict central values for $\Gamma_{N^*
\to \Lambda^*K}$ in the vicinity of 15 MeV for the current PDG value
of $M_{N^*}$. On the other hand, the width quoted in
\cite{simonprd58} leads to\footnote{The width is rather insensitive
to $g_2$ because its contribution is suppressed by the $K^+$ meson
c.m. three-momentum.} $|g_1|=1.25$, in good agreement with our
fitted value. The analysis carried out in \cite{Nam:2012ui} for the
$\Lambda(1520)$ electroproduction reaction off the proton uses this
value for $g_1$, fixes $g_2$ to 0 and, as mentioned above, uses the
helicity amplitudes given in \cite{Awaji:1981zj}. Thus, the set of
$N^*$ couplings used in \cite{Nam:2012ui}, where the important role
played by the $D_{13}$ $N^*(2120)$ resonance is also highlighted,
turns out to be similar to that found in this work.

The differential $d\sigma/d(\cos\theta_{\rm c.m.})$ distributions
from the combined fit (fit II) to the CLAS~\cite{Moriya:2013hwg} and
LEPS~\cite{leps2} data are shown in Figs.~\ref{dcs-clas} and
\ref{dcs-leps}, and compared with the experimental data. Only
statistical errors are displayed in these two figures. The
contributions from different mechanisms of the model are shown
separately.

In the first of these two figures, the differential cross sections
as a function of $\cos\theta_{\rm c.m.}$, for different $\gamma p$
invariant mass intervals, are displayed and contrasted with the
recent CLAS measurements. Near threshold the CLAS cross section is
fairly flat. In the highest energy bin the cross section is quite
forward peaked, with a hint of plateauing toward the most forward
angles. Also evident is that the cross section flattens or even
rises slightly toward large angles. We find an overall good
description of the data, both at forward and at backward $K^+$
angles and for the whole range of measured $\gamma p$ invariant
masses, $W$. We see that as $W$ increases, the contribution of the
$u-$channel $\Lambda (1115)$ pole term produces an enhancement at
backward angles, and it becomes more and more relevant. Indeed, it
turns out to be essential above $W \ge 2.35$ GeV and
$\cos\theta_{\rm c.m.} \le -0.5$. The present model provides a
better description of the recent CLAS data than that obtained within
the schemes of Refs.~\cite{nam3} and \cite{hejun}, whose predictions
are reported in \cite{Moriya:2013hwg}. The major improvement can be
appreciated at backward angles, as it is mostly attributable to the
$u-$channel $\Lambda (1115)$ pole mechanism.

For comparison, we also display in Fig.~\ref{dcs-clas} the
contribution from the $u-$channel $\Sigma^0 (1193)$ term, which is
not included in fit II. The $\Sigma^0 (1193)$ amplitude takes the
same form as that of the $\Lambda (1115)$ mechanism
(Eq.~(\ref{eq:au})), with the obvious replacements of coupling
constants and  hyperon masses. For the electromagnetic ones, we have
assumed $U-$spin invariance ($h_{1,2}^\Sigma=-\sqrt{3}h_{1,2}$),
while we have taken $g_{K^+p\Sigma^0} = 3.2$, as deduced from
leading order vector chiral symmetry (Eq.~(\ref{eq:chiral})). As
anticipated, the $u-$channel $\Sigma^0$ contribution turns out to be
a small fraction of the $\Lambda$ one, and can be safely neglected.

In Fig.~\ref{dcs-leps}, the differential cross section deduced from
the results of the nine-parameter fit II, as a function of the LAB
frame photon energy and for different forward c.m. $K^+$ angles, is
shown and compared both to CLAS~\cite{Moriya:2013hwg} and to
LEPS~\cite{leps2} data. In this figure, dashed (blue) and dotted
(green) lines show the contributions from the
background\footnote{The contribution of the $u-$channel $\Lambda
(1115)$ pole mechanism for these forward angles is negligible.} and
$N^*$ resonance terms, respectively, while the solid (red) lines
display the full result. We see that the bump structure in the
differential cross section at forward $K^+$ angles is fairly well
described thanks to the significant contribution from the $N^*$
resonance in the $s-$channel. Indeed, these results are similar to
those already reported and discussed in Ref.~\cite{xiejuan}, and the
description of these forward LEPS data achieved here is also
comparable to that exhibited in Ref.~\cite{xiejuan}. The CLAS data
points shown in Fig.~\ref{dcs-leps} were obtained from the
appropriate CLAS cross sections displayed in Fig.~\ref{dcs-clas},
relating $W$ to the LAB photon energy. Returning to the latter
figure, we see that our model underestimates the CLAS data for
values of $\cos\theta_{\rm c.m.}$ in the 0.5-0.75 interval, and in
particular at high energies, $W\ge 2.3$ GeV. These discrepancies
also show up in Figs.~\ref{dcs-leps}(a) and \ref{dcs-leps}(b), where
we can also appreciate that CLAS and LEPS data are only marginally
consistent. However, one should bear in mind that systematic errors
are not displayed in these plots, and they will make these
differences less meaningful.

We have also performed several best fits, where either one or both
of the $s-$channel $N^*(2120)$ resonance and the $u-$channel
$\Lambda (1115)$ pole terms have been switched off.  The
corresponding $\chi^2/dof$ are also compiled in
Table~\ref{tab:nstar}, which turn out to be larger, and in most
cases, unacceptable on statistical grounds.\footnote{Note that the
quantitative effect on $\chi^2/dof$ of disconnecting the $u-$channel
$\Lambda(1115)$ contribution is not big. This is because the forward
angle LEPS data are rather insensitive to this mechanism, as well as
the low energy CLAS cross section. As commented in the discussion of
Fig.~\ref{dcs-clas}, only the backward high energy CLAS cross
sections are being effected by this term, but there the qualitative
effect is important. Despite the limited number of data points,
which in addition suffer from large statistical fluctuations, the
improvement on the total $\chi^2/dof$ is still significant (3.0 vs
2.5).}

\begin{figure}
\begin{center}
\includegraphics[scale=1.]{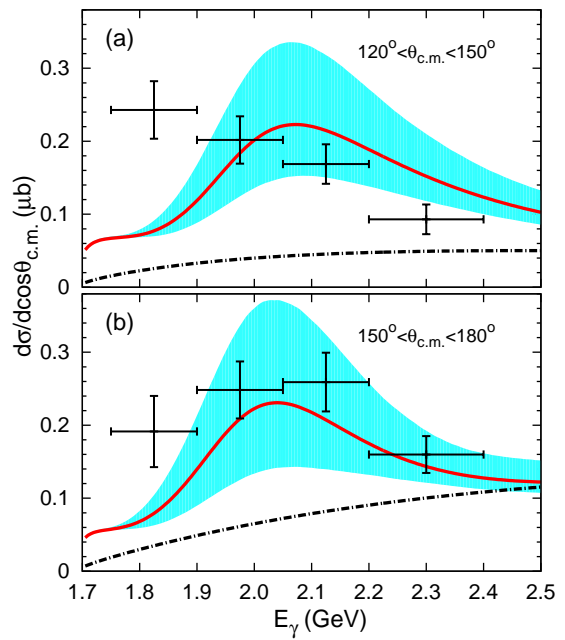}
\caption{(Color online) Fit II $\gamma p \to K^+ \Lambda (1520)$
differential  cross section as a function of the LAB frame photon
energy for backward $K^+$ angles [solid (red) lines]. We display the
average  $\left \langle d\sigma/d(\cos\theta_{\rm c.m.})
\right\rangle$ distribution (Eq.~(\ref{eq:avedcs})) from the full
model and compare it with LEPS data from Ref.~\cite{leps1}. (a, b)
The 68\% CL bands inherited from the Gaussian correlated statistical
errors of the fit II parameters. Finally, the dash-dotted (black)
curves show the $u-$channel $\Lambda (1115)$ contribution. }
\label{dcs-deg}%
\end{center}
\end{figure}

In Fig.~\ref{dcs-deg}, we compare the fit II predicted differential
cross sections for large $K^+$ angles with the experimental data
from Ref.~\cite{leps1}. There, events were accumulated for two
angular intervals $\theta_{\rm c.m.}$ =$(120 - 150)^0$ and
$\theta_{\rm c.m.}$ =$(150 - 180)^0$, with the photon energy varying
in the region $1.9\leq E_\gamma \leq 2.4$ GeV. Since the angular
intervals are quite wide, we have partially integrated the
differential cross section, and evaluated
\begin{equation}
\left\langle  \frac{d\sigma}{d(\cos\theta_{\rm c.m.})}  \right\rangle  =
\frac{\int^{\cos\theta_{\rm c.m.}^{\rm dw}}_{\cos\theta_{\rm
c.m.}^{\rm up}}
  \frac{d\sigma}{d(\cos\theta_{\rm c.m.})} d(\cos\theta_\text
{c.m.})}{\int^{\cos\theta_{\rm c.m.}^{\rm dw}}_{\cos\theta_{\rm
c.m.}^{\rm up}} d(\cos\theta_\text {c.m.})}, \label{eq:avedcs}
\end{equation}
taking ($\theta_{\rm c.m.}^{\rm dw}$, $\theta_{\rm c.m.}^{\rm up}$)
to be ($120^0$, $150^0$) or ($150^0$, $180^0$), respectively. In
Fig.~\ref{dcs-deg}, the shaded regions account for the uncertainties
inherited from those affecting the parameters compiled in
Table~\ref{tab:nstar}. They represent 68\% confidence-level (CL)
bands and were obtained using a Monte Carlo simulation. As shown in
Figs.~\ref{dcs-deg}(a) and \ref{dcs-deg}(b), the present model
provides a fair description of these backward $K^+$ angular data.
This is in sharp contrast to the results of our previous
work~\cite{xiejuan} (see Fig.~4 in Ref.~\cite{xiejuan}), where the
$u-$channel $\Lambda (1115)$ mechanism was not considered. Indeed,
the latter contribution is also depicted in Fig.~\ref{dcs-deg}. We
see that it increases with the photon energy and that becomes quite
relevant for the most backward angles [Fig.~\ref{dcs-deg}(b)].

\begin{figure}
\begin{center}
\includegraphics[scale=1.]{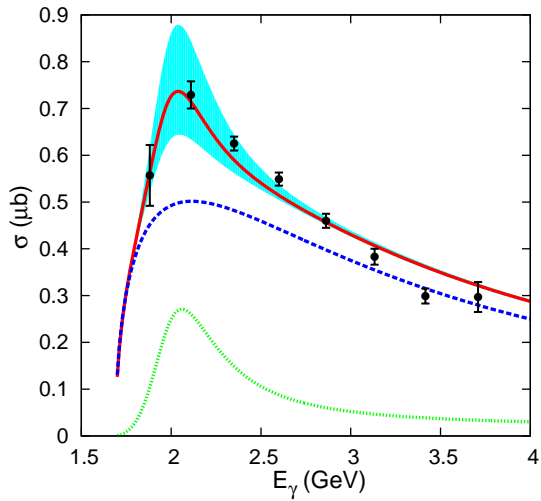}%
\caption{(Color online) Total $\gamma p \to K^+ \Lambda^*$ cross
section deduced from the results of fit II. We display it as a
function of the LAB photon energy and compare with the CLAS data
from Ref.~\cite{Moriya:2013hwg}. Dashed (blue) and dotted (green)
lines show the contributions from the background and $N^*$ resonance
terms, respectively, while the solid (red) line displays the results
from the full model. The shaded region accounts for the 68\% CL band
inherited from the Gaussian correlated statistical errors of the fit II parameters. } \label{Fig:tcs}%
\end{center}
\end{figure}

Finally, we have also calculated the total cross section of the
$\gamma p \to K^+ \Lambda(1520)$ reaction as a function of the
photon energy. The results are shown in Fig.~\ref{Fig:tcs} and
compared to the experimental data from CLAS.  We see that the model
provides an excellent description of the integrated CLAS cross
sections thanks to an important contribution from the
photo-excitation of the $N^*(2120)$ resonance and its subsequent
decay into a $\Lambda(1520)K^+$ pair. This mechanism seemed also to
be responsible for the bump structure in the LEPS differential cross
section at forward $K^+$ angles discussed in Fig.~\ref{dcs-leps}.
Thus, one can definitely take advantage of the apparently important
role played by this resonant mechanism in the LEPS and CLAS data to
better constrain some of the $N^*(2120)$ properties
(Table~\ref{tab:nstar}), starting from its mere existence.

It is noteworthy that the contribution from the $u-$channel
$\Lambda(1115)$ mechanism is very small in the integrated cross
section, as it is only significant for backward $K^+$ angles.
\section{Summary and Conclusions } \label{sec:conclusions}

We have carried out a new analysis of the $\gamma p \to \Lambda
(1520 )K^+$ reaction at low energies within an effective Lagrangian
approach and the isobar model. We have presented results from a
combined fit to the recent CLAS~\cite{Moriya:2013hwg} and
LEPS~\cite{leps2}, $d\sigma/d(\cos\theta_{\rm c.m.})$ data. Within
the scheme of Ref.~\cite{xiejuan}, and in addition to the contact,
$t-$channel $\bar K$ exchange, and $s-$channel nucleon  and
$N^*(2120)$ resonance pole contributions, we have also studied the
$u-$channel $\Lambda(1115)$ hyperon pole term. The latter mechanism
has been ignored in all previous
calculations~\cite{xiejuan,hejun,nam,toki} that relied on the very
forward $K^+$ angular LEPS data~\cite{leps2,leps1}, where its
contribution was expected to be small.

We have shown that when the contributions from the $N^*(2120)$
resonance and the $\Lambda(1115)$ are taken into account, both the
new CLAS and the previous LEPS data can be simultaneously described.
Actually, we find an overall good description of the data, both at
forward and at backward $K^+$ angles, and for the whole range of
measured $\gamma p$ invariant masses. The contribution of the
$u-$channel $\Lambda (1115)$ pole term produces an enhancement at
backward angles, and it becomes more and more relevant as the photon
energy increases, becoming essential above $W \ge 2.35$ GeV and
$\cos\theta_{\rm c.m.} \le -0.5$. On the other hand, the CLAS data
(see for instance Fig.~\ref{Fig:tcs}), clearly support the existence
of an odd parity $3/2$ wide nucleon resonance with a mass in the
region of 2.1 GeV, a width of around 200 MeV and a large partial
decay width into $\Lambda^*K$. The recent analysis carried out in
\cite{Nam:2012ui} of $\Lambda(1520)$ electroproduction off the
proton also concludes that the $D_{13}$ $N^*(2120)$ resonance plays
an important role. It is re-assuring that the $N^*$ couplings used
in \cite{Nam:2012ui} turn out to be quite similar to those
determined in this work. These characteristics could be easily
accommodated within the constituent quark model results of Capstick
and Roberts in Ref.~\cite{simonprd58}. Such resonance might be
identified with the two stars PDG $N^*(2120)$ state. This would
confirm previous claims~\cite{xiejuan,hejun} from the analysis of
the bump structure in the LEPS differential cross section at forward
$K^+$ angles discussed in Fig.~\ref{dcs-leps}, and contradict
previous negative claims made in \cite{nam2} and \cite{nam3}
regarding this point.

Following \cite{simonprd58}, besides the $N^*(2120)$ $D_{13}$
resonance, both of the weakly established states $N(2090)$ $S_{11}$
and $N(2200)$ $D_{15}$ [in the 2012 PDG review~\cite{pdg2012}, these
excited nucleons appear in the listings as the two-star $N(1895)$
and $N(2060)$, respectively], were also visible in $\Lambda(1520)K$
production reactions [see Fig. 6 of Ref.~\cite{simonprd58}]. If
these resonances were take into account, one might wonder whether
their contributions could affect the $N^*(2120)$ parameters in a
significant way. Fortunately, the work of S. Nam on
electroproduction of $\Lambda(1520)$ off the
nucleon~\cite{Nam:2012ui} sheds some light into this issue. There,
the $D_{13}(2120)$, $S_{11}(2090)$ and $D_{15}(2200)$ resonances
were considered, and it was found that the contributions of the two
latter ones are negligible and much smaller than that of the
$D_{13}(2120)$ state [see Figs. 6(a) and 11(a) in
Ref.~\cite{Nam:2012ui}].

On the other hand, the $N^*$ resonance parameters from the new fits
carried out in this work turn out to be in reasonable agreement with
those obtained in Ref.~\cite{xiejuan}, and the bulk of the
conclusions of that reference still hold. In particular, the sign
discrepancy (see Fig.~6 in \cite{xiejuan}) of the predictions of the
model for the polar-angle average photon-beam asymmetry, as a
function of $E_\gamma$, with the SPring-8 LEPS data of
Ref.~\cite{leps2} still persists.

In summary, we conclude that the associated strangeness production
reaction $\gamma p \to K^+ \Lambda(1520)$ is an adequate tool to
study the properties of the $N^*(2120)$ resonance, and provides
strong hints of its existence. This would corroborate the
theoretical expectations of the chiral inspired
unitary~\cite{Oset:2009vf,Gamermann:2011mq} and constituent
quark~\cite{simonprd58} models, and would make more plausible the
analysis of the $\gamma p \to K^0\Sigma^+$ CBELSA/TAPS data carried
out in \cite{Ramos:2013wua}, where the existence of a $J^P=3/2^-$
nucleon excited state around 2 GeV has also been claimed. In
addition, the study of the $\gamma p \to K^+ \Lambda(1520)$ reaction
also sheds light on the structure of
$\Lambda(1520)$~\cite{Roca:2006sz,Gamermann:2011mq}, and  its
properties such as the $\bar K^* N \Lambda^*$~\cite{toki} and
$K\Lambda^*N^*$ vertices (this work and Ref.~\cite{xiejuan}) and its
radiative $\Lambda^*\to \Lambda\gamma$ decay [$h_1$ and $h_2$
magnetic couplings, Eq.~(\ref{eq:eqgamaN}), determined in this
work].

\section*{Acknowledgments}

We warmly thank K.~Moriya for sending us the CLAS experimental data
files, and M.J. Vicente-Vacas for helpful discussions. This work was
partly supported by DGI and FEDER funds, under Contract No.
FIS2011-28853-C02-01 and FIS2011-28853-C02-02, the Spanish
Ingenio-Consolider 2010 Program CPAN (CSD2007-00042), Generalitat
Valenciana under contract PROMETEO/2009/0090 and by the National
Natural Science Foundation of China under grant 11105126. We
acknowledge the support of the European Community-Research
Infrastructure Integrating Activity "Study of Strongly Interacting
Matter" (HadronPhysics3, Grant Agreement No. 283286 ) under the
Seventh Framework Programme of EU. The work was supported in part by
DFG (SFB/TR 16, "Subnuclear Structure of Matter").

\end{document}